\documentclass[prb,twocolumn,showpacs,amsmath,amssymb,superscriptaddress]{revtex4}
\usepackage{graphicx}
\usepackage{dcolumn}
\usepackage{bm}
\usepackage{epsfig}
\begin{document}


\title{Spin freezing and dynamics in Ca$_{3}$Co$_{2-x}$Mn$_{x}$O$_{6}$ ($x \approx 0.95$)
investigated with implanted muons: disorder in the 
anisotropic next-nearest neighbor Ising model}

\author{T. Lancaster}
\email{t.lancaster1@physics.ox.ac.uk}
\author{S.J. Blundell}
\author{P.J. Baker}
\author{H.J. Lewtas}
\author{W. Hayes}
\affiliation{
Oxford University Department of Physics, Clarendon Laboratory,  Parks
Road, Oxford, OX1 3PU, UK}
\author{F. L. Pratt}
\affiliation{
ISIS Facility, Rutherford Appleton Laboratory, Chilton, Oxfordshire OX11 0QX, UK}
\author{H. T. Yi}
\author{S.-W. Cheong}
\affiliation{Rutgers Center for Emergent Materials and Department 
of Physics \& Astronomy, 
136 Frelinghuysen Road, Piscataway, New Jersey, 08854, USA}
\date{\today}

\begin{abstract}
We present a muon-spin relaxation investigation
of the Ising chain magnet Ca$_{3}$Co$_{2-x}$Mn$_{x}$O$_{6}$ ($x \approx 0.95$).
We find dynamic spin fluctuations persisting down to the lowest measured temperature of 
1.6~K. The previously observed transition at around 
$T \approx 18$~K is interpreted as a subtle change in dynamics for a 
minority of the spins coupling to the muon that we interpret as spins locking into clusters.
The dynamics of this fraction of spins freeze below a temperature $T_{\mathrm{SF}}\approx 8$~K, 
while a majority of spins continue to fluctuate. An explanation of the low temperature behavior is
suggested in terms of the predictions of the anisotropic next-nearest-neighbor Ising model.
\end{abstract}

\pacs{75.50.Ee, 75.50.Lk, 76.75.+i}
\maketitle

The magnetic chain multiferroic Ca$_{3}$Co$_{2-x}$Mn$_{x}$O$_{6}$ ($x \approx 1$) 
has been the subject of considerable recent investigation \cite{choi,jo,wu,zhang}. 
This material is based on the Ising spin chain magnet Ca$_{3}$Co$_{2}$O$_{6}$, with
(close to) half of the cobalt ions replaced with manganese \cite{zubkov,rayaprol}.
The observation of up-up-down-down ($\uparrow \uparrow \downarrow \downarrow$) order in this
system has led to the proposal that, at low temperatures, Ca$_{3}$Co$_{2-x}$Mn$_{x}$O$_{6}$  
may be described by the anisotropic next-nearest-neighbor Ising (ANNNI) model \cite{choi}. 
This model \cite{bakreview,bak,fisher,yeomans} describes Ising spins
on a three-dimensional lattice in which, along one direction, there is 
nearest-neighbor ferromagnetic exchange ($J_{\mathrm{FM}}$) and next-nearest-neighbor
antiferromagnetic exchange ($J_{\mathrm{AFM}}$). 
For $|J_{\mathrm{AFM}}/J_{\mathrm{FM}}|>1/2$ the ground state magnetic order is of the 
$\uparrow  \uparrow \downarrow \downarrow$ type. As temperature is increased
from $T=0$ the magnetic behavior is determined by the existence of 
domain wall solitons which separate regions with different commensurate AFM spin arrangements\cite{bak,fisher}. 
Although a continuum description of
the ANNNI model predicts an infinity of high order commensurate AFM phases (known as the devil's staircase)
a description in terms of a discrete Hamiltonian shows the possibility of metastable states of randomly 
pinned solitons. In magnetic systems, these so-called ``chaotic states'' are expected to lead to frozen-in disorder
or spin glass-like behavior \cite{bakreview}. 
Here we present an investigation of the 
low temperature static and dynamic magnetism in Ca$_{3}$Co$_{2-x}$Mn$_{x}$O$_{6}$ ($x \approx 1$)
that we have observed at a local level using muon-spin relaxation ($\mu^{+}$SR).  
We find that the low temperature magnetic 
state of Ca$_{3}$Co$_{2-x}$Mn$_{x}$O$_{6}$ is reached through a complex 
freezing out of dynamic processes and we conjecture that
the existence of chaotic states provides an explanation for the disordered magnetism and persistent
dynamics that we observe at low temperature. 

Ca$_{3}$Co$_{2-x}$Mn$_{x}$O$_{6}$ is formed from chains of magnetic ions arranged along the $c$-axis  
 in alternating oxygen cages of face-shared trigonal prisms and octahedra. 
Mn$^{4+}$ ions preferentially occupy the octahedral sites while the trigonal prisms are
occupied by Co$^{2+}$ ions. The magnetic chains form a triangular lattice in the $ab$ plane
 separated by Ca$^{2+}$ ions. 
While it is agreed that 3d$^{3}$ Mn$^{4+}$ is in the $S=3/2$ 
high-spin configuration, the spin state of 3d$^{7}$ Co$^{2+}$ has been questioned. Although 
fits to magnetic neutron diffraction data \cite{choi} suggest a low spin $S=1/2$
state, electronic structure 
calculations  \cite{wu,zhang} and x-ray absorption spectroscopy \cite{wu}
favor  the high spin $S=3/2$ state.
The neutron diffraction and magnetic susceptibility measurements \cite{choi}
indicate that below $T_{\mathrm{B}} \approx 18$~K spins align along
the $c$-axis, adopting the $\uparrow \uparrow \downarrow \downarrow$ configuration
with Mn and Co ordered moments of 0.66$\mu_{\mathrm{B}}$ and 1.93$\mu_{\mathrm{B}}$ respectively. The considerable width of the magnetic Bragg peaks suggests that this is not a state of
true long-range order (LRO), but rather represents the locking-in of spins into
finite sized domains. 
Taken with the Ising-like character of the magnetic ions, the observation of 
$\uparrow \uparrow \downarrow \downarrow$ order is suggestive that this material can be described as
a realization of the ANNNI model, or its extension \cite{kim} to the case of chains of 
unequal Ising spins.
It is also notable that the inversion symmetry breaking of $\uparrow  \uparrow \downarrow \downarrow$
order, along with the alternating charge order, results in magnetism driven ferroelectricity
in this material \cite{choi}.

\begin{figure*}
\begin{center}
\epsfig{file=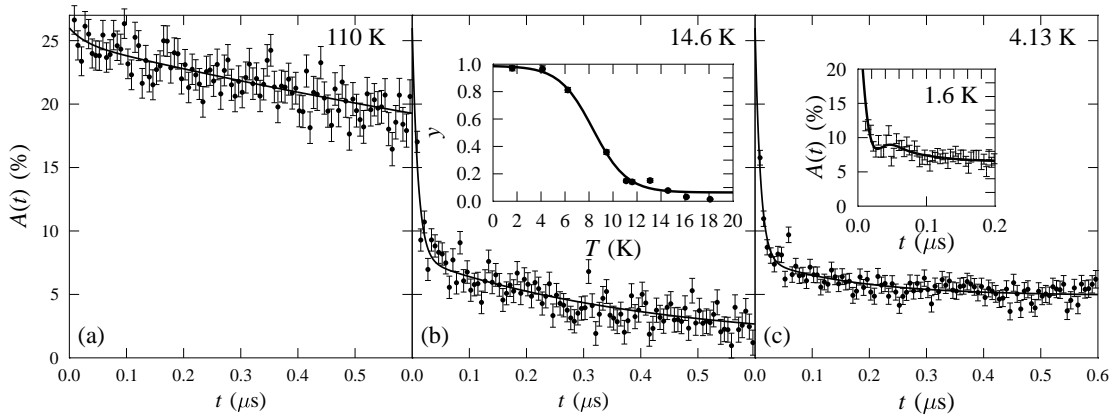,width=15cm}
\caption{ZF $\mu^{+}$SR spectra measured at (a) $T=110$~K, 
(b) 14.6~K and (c) 4.13~K. Solid lines are fits to Eq.~(\ref{fiteq}). 
{\it Inset} to (b): the parameter $y$ 
from Eq.~(\ref{fiteq}) shows evidence of static freezing around
$T \approx 8$~K. 
{\it Inset} to (c): Spectra measured at 1.6~K, showing a small, heavily damped
oscillation. The solid line is a guide to the eye. 
\label{data}}
\end{center}
\end{figure*}

$\mu^{+}$SR has proven useful in elucidating the static 
and dynamic properties of Ising systems \cite{le}
including the parent compound Ca$_{3}$Co$_{2}$O$_{6}$ \cite{takeshita,sugiyama}.
In a $\mu^{+}$SR experiment \cite{steve}, spin-polarized
positive muons are stopped in a target sample, where the muon usually
occupies an interstitial position in the crystal.
The observed property in the experiment is the time evolution of the
muon spin polarization, the behavior of which depends on the
local magnetic field at
the muon site, and which is proportional to the
positron asymmetry function $A(t)$. 
Measurements were made 
at the Swiss Muon Source (S$\mu$S) using the GPS instrument.
A polycrystalline sample of Ca$_{3}$Co$_{2-x}$Mn$_{x}$O$_{6}$ with $x=0.95$ (similar composition to
Ref.~\onlinecite{choi}), was packed in 
a Ag foil packet and mounted on a Ag plate in a helium cryostat.

Spectra measured for Ca$_{3}$Co$_{1.05}$Mn$_{0.95}$O$_{6}$ are shown in Fig.~\ref{data}. 
For temperatures $T>10$~K [Fig.~\ref{data}(a) and (b)] we observe purely relaxing asymmetry spectra.
We do not observe oscillations in the asymmetry, which usually signal the presence of long-range
magnetic order (LRO). 
For $T<8$~K the form of the spectra alters, most notably with an increase in the non-relaxing
baseline of the asymmetry. There is also a small undulation in the asymmetry  
that emerges at early times in high
statistics spectra measured at the lowest temperatures 
[inset Fig.\ref{data}(c)].
This feature, which may represent a low amplitude,  heavily damped oscillation 
displayed little temperature dependence and was too indistinct to be fitted systematically. 
Instead, the spectra are best modelled over the entire temperature range
by a sum of asymmetry functions, one with a large relaxation rate $\lambda_{1}$ and the other with
 a smaller rate $\lambda_{2}$:
\begin{eqnarray}
\label{fiteq}
A(t) &=&  A_{\mathrm{bg}} + A_{0} \left( 
p e^{-\lambda_{1} t} \right. \\ \nonumber
& & \left.
+ (1-p) \left[
\frac{y}{3} + \left(1-\frac{y}{3} \right) e^{-\lambda_{2} t} \right] \right),
\end{eqnarray}
where $A_{0}$ represents the amplitude of the
signal arising from the sample and 
the term  $A_{\mathrm{bg}}$ represents a temperature independent, nonrelaxing background 
from those muons that
stop in the sample holder or cryostat tails.
The parameter $y=0$ above $\approx 10$~K but  takes nonzero values at low temperatures (see below). 
The amplitude $p$ was found to be $p=0.75$ across the entire
measured temperature regime.
Exponential relaxation is often expected in cases where dynamic fluctuations in
the local magnetic field at the muon site represent the dominant relaxation 
process \cite{hayano} and it is likely that this mechanism is at work in
Ca$_{3}$Co$_{2-x}$Mn$_{x}$O$_{6}$. 
Fast field fluctuations lead to relaxation rates that vary as 
$\lambda_{i} \propto \gamma_{\mu}^{2} \langle B_{i}^{2} \rangle \tau$, where $\gamma_{\mu}$ is
the muon gyromagnetic ratio, $B_{i}$ is the local magnetic field at the $i$th muon site and $\tau$
is a fluctuation rate.
In the present case, the interpretation of dynamic fluctuations is supported by the observation that 
applied longitudinal magnetic fields of up to $0.6$~T 
do not decouple the relaxation,
as would be expected for relaxation from static field 
inhomogeneities \cite{decouplingnote}.

The coexistence of two relaxation rates, $\lambda_{1}$ and $\lambda_{2}$, implies
the existence of two classes of spatially separate muon sites in Ca$_{3}$Co$_{2-x}$Mn$_{x}$O$_{6}$. 
In general, these sites might differ in the width of the field distribution $\langle B^{2} \rangle$, 
or the fluctuation time at each site may be different. (It is not
possible to have a single class of site with two correlation times giving rise to 
two different relaxation rates; the shorter time will always dominate, giving the smaller
relaxation rate \cite{heffner}.)
The first class  of muon site (with amplitude $p$ and relaxation rate $\lambda_{1}$) 
accounts for $\approx 75$\% of the muon sites, with the second, with amplitude $(1-p)$,
 accounting for the remaining $\approx 25$\% 
of muon sites.
As there is no evidence (from our measurments or previous work \cite{choi})
for phase separation in this system, it is probable that both classes
arise from the intrinsic behavior of the bulk of the material. Possibilities for this 
include the coupling of each class of site preferentially to one or other of the magnetic cations 
(i.e.\ one class sensitive to 
fields arising from Co$^{2+}$ and the other sensitive to Mn$^{4+}$), as is the case in 
of $X_{3}$V$_{2}$O$_{8}$ ($X$=Ni,Co) \cite{lancaster1}, or that
the classes of muon site are sensitive to different components of the same spin distribution
as in GeNi$_{2}$O$_{4}$ \cite{lancaster2}. In the latter case, a system
with local site anisotropy might, for example, give rise to relaxation
 timescales for longitudinal and transverse fluctuations that could be quite
 different. If different muons sites were selectively sensitive to
 longitudinal or transverse components, dynamics with two timescales may arise
from the same spin site.
 
\begin{figure}
\begin{center}
\epsfig{file=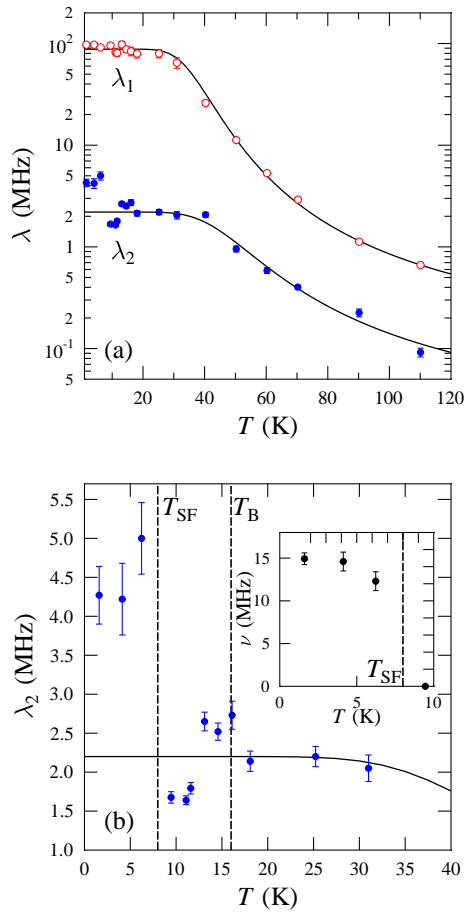,width=7.cm}
\caption{(a) Evolution of the relaxation rates $\lambda_{1}$ (open circles) and 
$\lambda_{2}$ (closed circles) as a function of temperature in ZF.
(b) Detail from (a) showing departure of $\lambda_{2}$ from the trend around the 
 transition at $T_{\mathrm{B}}$ and the spin freezing transition
$T_{\mathrm{SF}}$. 
Inset: Temperature evolution of the frequency of the small oscillatory feature seen below 8 K at 
early times.
\label{fit}}
\end{center}
\end{figure}

The temperature dependence of $\lambda_{1}$ and 
$\lambda_{2}$ resulting from fits of the data to Eq.~(\ref{fiteq})
are shown in Fig.~\ref{fit}.
Both relaxation rates show the same 
trend of behavior with their magnitude
increasing as the temperature is reduced, followed by saturation of the relaxation rate
at a constant value below $\approx 30$~K. This trend is sometimes seen in the $\mu^{+}$SR of
complex systems with low temperature dynamics \cite{tom,psd} and its origin is not completely understood. 
It can arise due to the existence of two relaxation
channels, one of which is strongly $T$ dependent with a correlation time $\tau_{\mathrm{s}}$
while the other shows little variation with $T$ and has a correlation time $\tau_{\mathrm{w}}$.
As noted above, in the presence of two competing relaxation processes, that with the shorter correlation
time wins out, giving the smaller relaxation rate. 
At high temperature, therefore, we have $\tau_{\mathrm{s}}(T) \ll \tau_{\mathrm{w}}$ which results in a
strongly $T$ dependent relaxation which we can fit phenomenologically by $\lambda_{i} = C_{i}  \exp (U_{i} / T)$. 
At low temperatures, where $\tau_{\mathrm{w}} \ll \tau_{\mathrm{s}}(T)$, 
we have $\lambda_{i} \sim \lambda^{\mathrm{sat}}_{i}$, 
resulting in a phenomenological fitting function 
$1/ \lambda_{i}(T) = 1/\lambda_{i}^{\mathrm{sat}} + 1/[C_{i} \exp (U_{i}/T)]$. 
Fitting this function to our data allows us to parameterize the relaxation rates with 
$\lambda^{\mathrm{sat}}_{1}= 88(3)$~MHz, $U_{1} \approx 275(5)$~K for the fast relaxation
and $\lambda^{\mathrm{sat}}_{2}=2.2(2)$~MHz, $U_{2} \approx 270(20)$~K for the slowly relaxing component. 
Given the similarity between $U_{1}$ and $U_{2}$,
it is likely that similar relaxation processes are at work at both sites, sharing similar correlation
times. This would imply that the 
widths of the magnetic field distributions differ for the two classes of site and are in the ratio 
$(\langle B^{2}_{1}\rangle/ \langle B^{2}_{2} \rangle)^{1/2}  \approx 
(\lambda^{\mathrm{sat}}_{1}/\lambda^{\mathrm{sat}}_{2})^{1/2} \approx 6$. 
We note further that, with the large activation
energies involved, it is unlikely that the temperature dependent behavior is related to
dynamics predicted by the ANNNI model. This is not surprising, since models of this sort
provide a low energy description of real magnetic systems.
Rather, it is likely that the observed behavior 
arises due to the single ion anisotropies of the magnetic ions. 
For free Co$^{2+}$, for example, the spin orbit coupling constant \cite{low} is $-270$~K, 
which resembles our energy scale $U$.
This is suggestive that the energy scale for the temperature dependent contribution to the relaxation
rates is set by fluctuations between spin-orbit split components of the low-lying
electronic states of the magnetic Co$^{2+}$ ions. It should be expected that Mn$^{4+}$ will 
affect the magnetic field distribution at the muon sites making the true situation more complex.

In addition to this general trend reflecting the dynamics of the system, 
the smaller relaxing component, with relaxation rate $\lambda_{2}$, 
shows additional features at low temperature (although it is not obvious why these features are 
not seen in $\lambda_{1}$).
For the material in this regime, where the ANNNI model provides a good effective description
of the magnetic behavior, the spin fluctuations will include excitations involving
the domain wall solitons described by the ANNNI model \cite{bakreview}, along with diffusive modes resulting 
from coupling of the Ising spins to non magnetic degrees of freedom. 
At around $T_{\mathrm{B}} \approx 18$~K
we see a small peak in the relaxation rate $\lambda_{2}$
which occurs near the temperature below which
(broadened) magnetic Bragg peaks appear in neutron diffraction \cite{choi}. 
The absence of a change at $T_{\mathrm{B}}$ in the form of the signal or its 
amplitudes indicates that this is not a transition to LRO or to a static magnetic state. 
Rather, it is more likely to represent the freezing out of one or more relaxation processes, 
likely to involve the free motion of domain walls, leading to the
formation of poorly correlated clusters of ordered spins.
This is consistent with the observation of broadened peaks in the neutron diffraction \cite{choi}
signalling that the order that gives rise to the magnetic Bragg peaks is not truly long range. 

The more significant change in the slowly relaxing component of the muon asymmetry occurs below 8~K. 
Here the parameter $y$ in Eq.~(\ref{fiteq}) increases on cooling from $y=0$ to $y=1$ at the
lowest measured temperatures, as shown in Fig.~\ref{data}, 
indicating a transition to a static local field distribution at low $T$. 
Such a distribution, whether ordered or disordered, will only
dephase those muon-spin components that lie perpendicular to the initial muon-spin polarization
direction, expected to be 2/3 of the total polarization for a powder sample, leaving the other 
1/3 polarized. Although the magnetic field distribution
 experienced by these muons below $T \approx 8$~K is static on the muon timescale,  it is unlikely that
the system locks into true LRO. 
Instead of the oscillations that would usually be observed in the presence of LRO, we
see only a very small minimum in the asymmetry at early times [inset Fig.~\ref{data}(c)] in the minority
component of our spectra measured below 8~K.
This may originate from Kubo-Toyabe-like relaxation \cite{hayano} typical of a static ensemble of 
disordered moments
or may be a highly damped oscillation due to static order occurring over only a short length scale.
Assuming the latter interpretation, the inset to Fig.~\ref{fit}(b) shows the frequency 
$\nu(= \gamma_{\mu} B_{i}/2 \pi)$ of this feature \cite{footnote}. 
The static disordered magnetism observed in our experiments, 
taken together with the peak measured in the $\chi''$ susceptibility 
and broad maximum in the heat capacity
observed around 8~K \cite{choi}, points to a freezing of domain walls
to form spin clusters, resulting in a static disordered or glassy magnetic system
\cite{spinglass}. There are also, of course, still significant slow 
dynamic fluctuations in the system, 
measured by the majority asymmetry component with amplitude $p$. 

The observed disorder may be explained in terms of the predictions of the ANNNI model \cite{bakreview}. 
Although the
devil's staircase of commensurate spin structures is predicted from a continuum treatment of the
model, a treatment based on a discrete Hamiltonian predicts the existence of an array of ``chaotic''
states at a higher energy than the devil's staircase, comprising a
random array of domain wall solitons and antisolitons. Such states are metastable, but
are separated from the true
Devil's staircase of stable states by relatively high energy barriers, making it impossible for a
system to relax into the devil's staircase states in a finite amount of time \cite{bakreview}. 
Instead of LRO, the chaotic state possesses an intrinsic randomness which means it may be 
described as spin glass-like. 
This picture therefore provides an explanation within the framework of the ANNNI model 
for the disordered magnetic state that we observe in CaCo$_{1.05}$Mn$_{0.95}$O$_{6}$. 
We note, however, that additionals factors not considered in this simple model
may contribute to the glassy character, including 
the possibility of clustering of the Mn$^{4+}$ and Co$^{2+}$ ions and the effect of 
complex interchain exchange interactions \cite{kiryukhin}.

Finally we note that
it is possible that the behavior observed in CaCo$_{1.05}$Mn$_{0.95}$O$_{6}$
reflects the existence of both series spin relaxation processes
(where the freezing of one relaxation process allows another to freeze at a lower temperature)
and also coexistent parallel relaxation processes which persist independently. Both of these
types of process have been advanced to explain dynamics in glassy systems \cite{palmer} 
and it is likely that they are important in a complex dynamic system such as Ca$_{3}$Co$_{1.05}$Mn$_{0.95}$O$_{6}$.

Part of this work was carried out at S$\mu$S, Paul Scherrer Institut, Villigen, CH. We are grateful to
H. Luetkens and A. Amato for experimental assistance
and to J.M. Yeomans for useful discussions. This work is supported by the EPSRC (UK).
Work at Rutgers was supported by the DOE under Grant No. DE-FG02-07ER46382.


\begin{thebibliography}{xx}
\bibitem{decouplingnote}
Decoupling implies $\lim_{t \rightarrow \infty}A(t)=A(0)$. 
The maximum available field using the GPS spectrometer is 0.6~T.
\bibitem{footnote}
A full analysis of this feature is difficult and requires the replacement of
$e^{-\lambda_{2}t}$ in Eq.(\ref{fiteq}) with $(1-a) e^{-\lambda_{2}t}+a e^{-\lambda_{3}t}\cos(2 \pi \nu t)$, where 
$a (\ll 1)$ is an amplitude and $\lambda_{3}$ a large relaxation rate. Since this approach does not 
alter our interpretation we do not persue it here.
\end{thebibliography}
\end{document}